\documentclass{siamltex}
\usepackage{amsmath}
\usepackage{graphicx}
\DeclareGraphicsExtensions{.jpg,.pdf,.png,.mps}

\newcommand{\bx}{{\bf x}}

\newcommand{\ce}{{\cal E}}

\begin{document}

\title{Modelling and Simulations of Multi-component Lipid
Membranes and Open Membranes via Diffusive Interface
Approaches\thanks{This research is supported in part by NSF-DMS
0409297 and NSF-ITR 0205232.}}

\author{Xiaoqiang Wang\thanks{Institute for
Mathematics and its Applications, University of Minnesota, 
Minneapolis, MN 55455.
 {wang@ima.umn.edu}}
  \and
Qiang Du\thanks{Department of Mathematics, Penn State
  University, University Park, PA 16802.
  {qdu@math.psu.edu}}
}

\date{Received: date / Revised version: date}
%
\maketitle


\begin{abstract} In this paper, phase field models are developed for
multi-component vesicle membranes with different lipid
compositions and membranes with  free boundary. These models are used
to simulate the deformation of membranes
under the elastic bending energy and the line tension energy with
prescribed volume and surface area constraints. By comparing our
numerical simulations with recent experiments, it is 
demonstrated that the phase field models can
capture the rich phenomena associated with the
membrane transformation, thus it
offers great functionality in the simulation and modeling of
multicomponent membranes. 
\end{abstract}

\section{Introduction}

Lipid vesicle membranes are ubiquitous in
biological systems.  Studies of vesicle self
assembly and shape transition, including bud formation
\cite{Li92,Li95,Se92} and vesicle fission \cite{DoKaNoSp93} are very
important in the understanding of cell functions. In recent
experimental studies, multi-component vesicles with different
lipid molecule compositions (and thus phases) have been shown to
display even more complex morphology involving rafts and
micro-domains \cite{BHW03}. There are strong evidences suggesting
that phase segregation and interaction contribute critically
to the membrane signaling, trafficking and sorting processes
\cite{BSWJ05}.  In the literature, the geometric and topological
structures of multi-component vesicles  have been theoretically
modeled by minimizing an energy with contributions of the bending
resistance, that is the elastic bending energy, and the line
tension at the interface between different components or the phase
boundary \cite{BeMc92,gg98,kgl01,Li02}. 
The elastic bending energy first studied
by Canham, Evans and Helfrich \cite{Ci98,Ci00,OuLiXi99} for a
single-phase membrane is defined as
\begin{equation}\label{melas}
E  =  \int_\Gamma \left(a_1 + a_2 (H-c_0)^2 + a_3 G \right) \, ds,
\end{equation}
where $H$ is the mean curvature of the membrane surface
$\Gamma$,
$c_0$ the spontaneous curvature,
$G$ the Gaussian curvature, $a_1$
the surface tension, $a_2$  the bending rigidity and $a_3$ the
stretching rigidity.

In recent experimental studies, it has been found that the bending
rigidity in the liquid-disordered phase differs from that in the
liquid-ordered phase in two-component membranes
\cite{BHW03,BSWJ05}. This can be attributed to, among other
things, that the two phases have different lipid compositions or
different concentrations of cholesterol molecules which serve as 
spacers between lipids.
Thus, in the generalized bending elasticity model
for two-component membranes, 
the bending rigidity $a_2$ is assumed to take values $k_1$ and $k_2$
respectively in two different components (phases) 
$\Gamma_1\subset \Gamma$ and $\Gamma_2\subset \Gamma$ (with 
$\Gamma=\Gamma_1\cup\Gamma_2$).
The common phase boundary between the two phases is denoted by
$\gamma_0=\Gamma_1\cap\Gamma_2$.
In general, the other parameters may also vary in different phases,
however, in this paper, we ignore the effect due to $a_1,a_3$ and
$c_0$ and concentrate only on the effect of the bending rigidity,
though the methodology can be easily extended to more general case.
In fact, the formulation we present here  works for two components having
possibly different spontaneous curvatures, but for simplicity
such curvatures are set to be zero in the numerical simulations.

With the two phases co-existing on the membrane, it is natural to
introduce a line tension on $\gamma_0$ to take into account the
interfacial energy between the individual components
\cite{BHW03,lip96,Se92}. Coupling with the bending
elastic energy, this leads to the following total energy
determining the two component membrane
\begin{equation}\label{metotal}
E  =  E_1 +E_2 +E_l =\sum_{i=1}^2\int_{\Gamma_i} k_i (H-c_i)^2 \,
ds +
 \int_{\gamma_0} \delta  \,dl\;,
\end{equation}
where $\delta$ is the line tension constant \cite{lip96}. Note that in general,
the line energy can also include the integral of a multiple of the
curvature square on $\gamma_0$ \cite{BSWJ05}.

The mathematical model that our study is based on is the minimization
of the total energy defined in (\ref{metotal}) for a two component membrane
with a prescribed total volume, and prescribed surface areas of
both components. In order to effectively model and simulate the
experimental findings on the exotic morphology of the
multi-component vesicles (mostly taken from \cite{BHW03}), 
we extend the recently developed phase field approach for the single 
component vesicles \cite{DuLiWa04} to the multi-component case, which 
avoids the tracking of the vesicle membrane by viewing the surface and 
phase boundary as the zero
level sets of phase field functions. The general phase field
framework has been used successfully in
many applications \cite{AMW98,BWBK02,CaCh92,Ch02}. For membrane deformation,
this approach has become increasing popular in
the research community in recent years. So far, its applications 
have mostly confined to the case of using a single
phase field function \cite{misbah,DuLiRoWa042,DuLiRyWa06,jiang00,MAV04},
albeit it is known that co-dimension two objects can be described
effectively by a pair of level-set or
phase field functions \cite{osher01,Ch02,OR02,Se99}. With
the introduction of a second phase field function, we demonstrate
that the new two-component phase field model is capable of
capturing rich complex morphological
changes experimentally observed in the two-component vesicle membranes. Moreover, 
this model can be very easily generalized to study the open membranes or membranes with
free boundary (see \cite{SaTaHo98} for experimental study and 
\cite{cgs02,TuOu03,TuOu04,umeda,yyn05} for analysis and computation).
This is based on the observation that  an open membrane
can be thought as a two component vesicles with one component
having zero bending rigidity. Further generalization is possible
for vesicles with three or more components.

The paper is organized as follows: in section 2, we present the
phase field formulation of the total energy (\ref{metotal}) and
address the approach of penalty formulation for the constraints.
In section 3, we first briefly discuss the discretization schemes
and some implementation issue. After presenting some convergence
tests to validate our method, we assemble a number of interesting
experiments to explore the shape transformations due to the
changes of different parameters. The numerical simulations are
compared with experimental findings including the merge and
splitting of different components. In section 5, we present the
phase field formulation for open membranes and some numerical
simulation results. We then make some conclusion remarks in
section 6. Some technical derivations are provided in the
appendix.

\section{A diffusive interface model}

We start by introducing a pair of phase field functions
$(\phi(\bx), \eta(\bx))$, defined on the physical (computational)
domain $\Omega$.

The function $\phi=\phi(\bx)$ is used so that the level set
$\{\bx: \phi{(\bx)} = 0\}$ gives the membrane $\Gamma$, while
$\{\bx: \phi{(\bx)} >0 \}$ represents the interior of the membrane
(denoted by $\Omega_i$) and $\{\bx: \phi{(\bx)} < 0 \}$ the
exterior (denoted by $\Omega_e$). In the phase field models of a
single component vesicle, this is the only phase field function
used \cite{DuLiWa03}.

Next, we take another closed surface $\Gamma_\bot$ defined on
domain $\Omega$ and being perpendicular to $\Gamma$, such that it is 
the zero level set $\{\bx: \eta{(\bx)} = 0\}$ of a phase field
function $\eta=\eta(x)$ in $\Omega$ with $\{\bx: \eta{(\bx)} >0 \}$ 
being the interior of $\Gamma_\bot$ 
and $\{\bx: \phi{(\bx)} < 0 \}$ the exterior.
 We thus take the part of $\Gamma$ in the interior of $\Gamma_\bot$ as
the first component $\Gamma_1$
and the remain part of $\Gamma$  (denoted by $\Gamma_2$)
makes up the second component. 
Note that there may be many choices to
select $\Gamma_\bot$, but we are mostly interested in 
the level set $\{\bx: \eta{(\bx)} = \phi{(\bx)} = 0\}$ which gives the
boundary between two components, with $\{\bx: \eta{(\bx)}>0
\mbox{ and } \phi(\bx) = 0 \}$ representing one component of the
membrane and $\{\bx: \eta{(\bx)} < 0 \mbox{ and } \phi(\bx) = 0\}$
the other component. 

In the phase field modelling, the functions
$\eta$ and $\phi$ are forced to be nearly constant valued except
in thin regions near the surfaces $\Gamma$ and $\Gamma_\bot$
respectively.  We use two small positive constant parameters
$\epsilon$ and $\xi$ to characterize the widths of the thin
regions (also called the diffusive interfaces). We note that a phase field
function (order parameter), like our $\eta$,  has been introduced
in \cite{jiang00,MAV04} to describe the phase segregation on the
membranes, but different from our phase field description of the
surface $\Gamma$, an explicit construction of the membrane surface
and a direct computation of the bending elastic energy are used there
instead of the phase field representation of the membrane surface.

Similar to \cite{DuLiRoWa042},
we have the phase field elastic bending energy defined by
\begin{equation} \label{eme}
E(\phi, \eta) = \int_{\Omega} \frac{k(\eta)}{2\epsilon}
\big(\epsilon\Delta\phi + ( \frac{1}{\epsilon} \phi
+c_0(\eta) \sqrt 2)(1-\phi^2)
\big)^2 \, dx,
\end{equation}
where we take a variable bending rigidity given by $k(\eta) = k +
c \tanh(\frac{\eta}{\xi})$, so that $k + c$ corresponds to the value of the
bending rigidity of one component and $k-c$ the other. 
Similarly, 
 $2 c_0(\eta) = (c_1+c_2) + (c_1-c_2) \tanh(\frac{\eta}{\xi})$, 
so that $c_1$ and $c_2$ correspond to the 
spontanenous curvatures in the two components respectively.
A few other functionals needed in our model are as follows:
\begin{equation} \label{eml}
L(\phi, \eta)
 = \int_\Omega \delta [\frac{\xi}{2}|\nabla\eta|^2
+ \frac{1}{4\xi}(\eta^2 -1)^2][\frac{\epsilon}{2}|\nabla\phi|^2
+ \frac{1}{4\epsilon}(\phi^2 -1)^2] \, dx,
\end{equation}
\begin{equation}
A(\phi) = \int_\Omega [ \frac{\epsilon}{2} |\nabla\phi|^2 +
\frac{1}{4\epsilon}(\phi^2 -1)^2 ] \, dx,
\end{equation}
\begin{equation}
A(\phi) = \int_\Omega \phi \, dx,
\end{equation}
\begin{equation}
D(\phi, \eta) = \int_\Omega \tanh(\frac{\eta}{\xi})
[\frac{\epsilon}{2}|\nabla\phi|^2 + \frac{1}{4\epsilon}(\phi^2
-1)^2] \, dx\;.
\end{equation}

To reveal the meaning of the above functionals, we follow similar
discussions in \cite{DuLiRoWa042} to assume an ansatz of the form
$\phi(\bx) \sim \tanh(d(\bx,\Gamma)/(\sqrt 2 \epsilon))$ and
$\eta(\bx) \sim \tanh(d(\bx,\Gamma_\bot)/(\sqrt 2 \xi))$ for the
phase field functions. Here $d$ denotes the signed distance
function. In this ansatz, we can check that as $\epsilon $ and
$ \xi$ tend to $0$, that is, in the sharp interface limit,
\begin{equation}
\label{eqsharp1}
E(\phi, \eta) \to \frac{2\sqrt{2}}{3}
\sum_i\int_{\Gamma_i} k_i (H-c_i)^2 \, ds\;.
\end{equation} More details are
given in the appendix, along with a brief derivation of the
folllowing asymptotic limits
\begin{equation}
\label{eqsharp2}
V(\phi) \to 2 |\Omega_i| - |\Omega|  \;,\quad A(\phi)\to \frac{2
\sqrt{ 2 }}{3}|\Gamma| \;,
\end{equation}
 and
\begin{equation}
\label{eqsharp3}
L(\phi, \eta) \to \frac{8}{9} \int_{\gamma_0} \delta
dl\;, \qquad D(\phi,\eta) \to \frac{2\sqrt{2}}{3}  ( |\Gamma_1|
-|\Gamma_2|)\; .
\end{equation}

To re-cap the discussion, the total energy in the phase field two-component 
model is
\begin{equation} \label{ebe}
\ce(\phi, \eta) = E(\phi, \eta) + L(\phi, \eta)\;,
\end{equation}
while the constraints are given by
\begin{equation}
V(\phi) = v_d,\quad A(\phi) = a_0,\quad D(\phi, \eta) =a_d,
\end{equation}
with $v_d$, $a_t$ and $a_d$ being the prescribed volume
difference (hence the interior volume is prescribed), 
the total surface area  and the area difference between
the two components (hence areas of both components are prescribed).

To maintain the consistency of the phase field model which is based
on $\phi$ and $\eta$ having the tanh profiles and the orthogonality
between $\Gamma $  and $\Gamma_\bot$,  
additional constraints are imposed. First of all,
the orthogonality constraint on the normal directions
of the two surfaces,
written in our phase field formulations, can be enforced by
$\nabla \phi \cdot \nabla \eta = 0$ on or near the phase boundary
$\{\bx: \phi(\bx) =\eta(\bx) = 0\}$. With $\phi$ and $\eta$
having tanh profiles,
their gradients become small away from their zero level sets, the
orthogonality constraint may thus be enforced everywhere by penalizing
\begin{equation}\label{epe}
N(\phi, \eta) = \int_\Omega \frac{\epsilon}{2}|\nabla \phi \cdot
\nabla \eta|^2 \, dx \;.
\end{equation}
Secondly, to better maintain the tanh profile of $\eta$,
especially for the case with a large line tension energy, we have
two options, one is  to add a small regularization term, much like the
bending elastic energy for $\phi$ but with a very small bending rigidity;
another option is to regularize through the following functional
\begin{equation}
P(\eta) = \int_\Omega \big(\frac{\xi}{2}|\nabla\eta|^2 -
\frac{1}{4\xi}(\eta^2 -1)^2\big)^2 \, dx \;,
\end{equation}
which also vanishes for any function $\eta$ with a tanh profile.

Summarizing the above, the variational phase field model
to describe the two-component vesicles in the energy minimizing state
is to minimize  the total energy $\ce(\phi, \eta) = E(\phi,
\eta) + L(\phi, \eta) $ with constraints $V(\phi) = \alpha_1$,
$A(\phi) = \alpha_2$, $D(\phi, \eta) = \alpha_3$ while $N(\phi,
\eta) $ and $P(\eta)$ remain small. So, 
by adding both the penalty and regularization terms, the vesicle 
surface and the two components
phase boundary are determined by a pair of phase functions
$(\phi,\eta)$ which minimizes the energy
\begin{eqnarray}\
\ce_M(\phi, \eta) &=& E(\phi, \eta) + L(\phi, \eta) +
\frac{1}{2}M_1(V(\phi) - v_d)^2 + \frac{1}{2}M_2(A(\phi) - a_0)^2
\nonumber\\&&+ \frac{1}{2}M_3(D(\phi, \eta) - a_d)^2 +
\frac{1}{2}M_4(N(\phi, \eta))^2 + \frac{1}{2}M_5(P(\eta))^2
\label{tengy}
\end{eqnarray}
where $\{M_i\}_{i=1}^3$ are penalty constants for the constraints
on the volume and surface areas while 
 $\{M_i\}_{i=4}^5$ are regularization constants for
maintaining better control on the phase field functions.

\section{Numerical Simulations of Two-Component Membranes}\label{sec:mns}
In this section, we compute the minimum of the phase field energy 
(\ref{tengy}) 
by adopting a  gradient flow approach which has been very effective for solving the
phase field model of single component vesicles \cite{DuLiWa03,DuLiWa06}.
The equations for the gradient flow are given by:
\begin{equation}\label{eq3d}
\phi_t= - \frac{\delta \ce_M}{\delta \phi},\quad
\eta_t= -\frac{\delta \ce_M}{\delta \eta}\;.
\end{equation}
The monotone decreasing of the energy $E_M$ is ensured for $t>0$.
For simplicity, we only consider the case where $c_1=c_2=0$, this allows us
to focus on examining 
how the variation in the bending rigidity alone affects
the vesicle shape deformation and the equilibrium configurations of
two-component membranes. The more general cases involving the
spontaneous curvatures are to be considered in the future.

\paragraph{Discretization and code development.}
We take the spatial computational domain as the box $\Omega=[-\pi,
\pi]^3$ and assume that membranes are  enclosed in the box.
Moreover, we choose to set $\xi = \epsilon$ in our numerical
simulations. For the spatial discretization of (\ref{eq3d}) in 
$\Omega$, a Fourier spectral method
is used. Due to the regularization effect of the finite
transition layer, for fixed $\epsilon$ and enough Fourier modes,
the spectral method is an efficient way to solve (\ref{eq3d})
with the help of FFT routines \cite{Ch02}. 
A couple of options are implemented for
the time discretization, such as an explicit forward Euler scheme
or a semi-implicit Euler scheme \cite{DuLiWa03}. The
time step $\Delta t$ is chosen to ensure the energy decay. For
most of our numerical experiments, although fully adjustable,
$\Delta t$ is kept in the range of $10^{-6}$ to $10^{-7}$. The
simulation codes are fully parallelized on both distributed memory
systems via MPI and shared memory systems via OPENMP to improve
its efficiency and functionality in conducting extensive three
dimensional simulations. 

\paragraph{Problem set up and initial profiles.}
We now discuss how we choose various parameters in the simulations.
Though in theory the gradient flow can be started from any pair of
initial phase field functions, a proper choice often speeds up the
evolution process and allows more efficient solution of the
equilibrium state. With the penalty
formulation,  a particular constraint can be simply removed
by setting the corresponding penalty constant zero. For example,
setting $M_2 = M_3= 0$ would eliminate the total area and area
difference constraints. This fact can be utilized to find good initial
phase field functions.

\begin{figure}[h]
\centerline{
\includegraphics[width=3.0in]{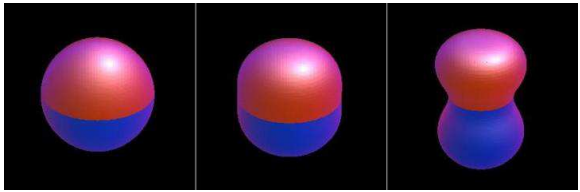}
} \caption{Line tension drives a two-component sphere to a gourd
shape.} \label{m01}
\end{figure}

For example, as illustrated in Figure.\ref{m01}, for a given $r>0$, we may
start from two special phase field functions as $\phi(\bx) =
\tanh(\frac{|x| - r}{\sqrt 2 \epsilon})$ and $\eta(\bx) = \tanh(\frac{z}{\sqrt 2 \epsilon})$ 
where $z$ is the third component of $\bx$. This provides
two hemispheres that represent the two components (colored in
$red$ and $blue$ respectively, or in gray-scale represented by
lighter and darker regions). Starting from this initial state,
and setting $M_1 = 0$ to eliminate the volume constraint,
the sphere gradually becomes more elliptical due to the presence
of line tension, then further transform to a gourd like shape.
We may stop at an intermediate shape and add back the volume
constraint. This would provide a variety of initial shapes
to be used in the simulations.

\paragraph{Convergence verification.}
For a particular numerical simulation, the quality of the numerical
result may be affected by the choice of computational domain, the
 parameter $\epsilon$ (the effective width of the
diffusive interface), the number of grid points, and the
choices of other parameters used in the simulation.
The parameter $\epsilon$ is generally taken to be a few
percentage points of the domain size to ensure a relatively sharp
interfacial region and the consistency with the sharp interface
description (the $\epsilon \to 0$ limit).  
The mesh size is normally taken to be several
times smaller than the width of the transition layer
to ensure adequate spatial resolution.
To ascertain the accuracy and robustness of our numerical algorithms
and the parameter selections, we here present results of some
numerical tests on the convergence and performance of our method.

\begin{figure}[t]
\centerline{
\includegraphics[width=2.0in]{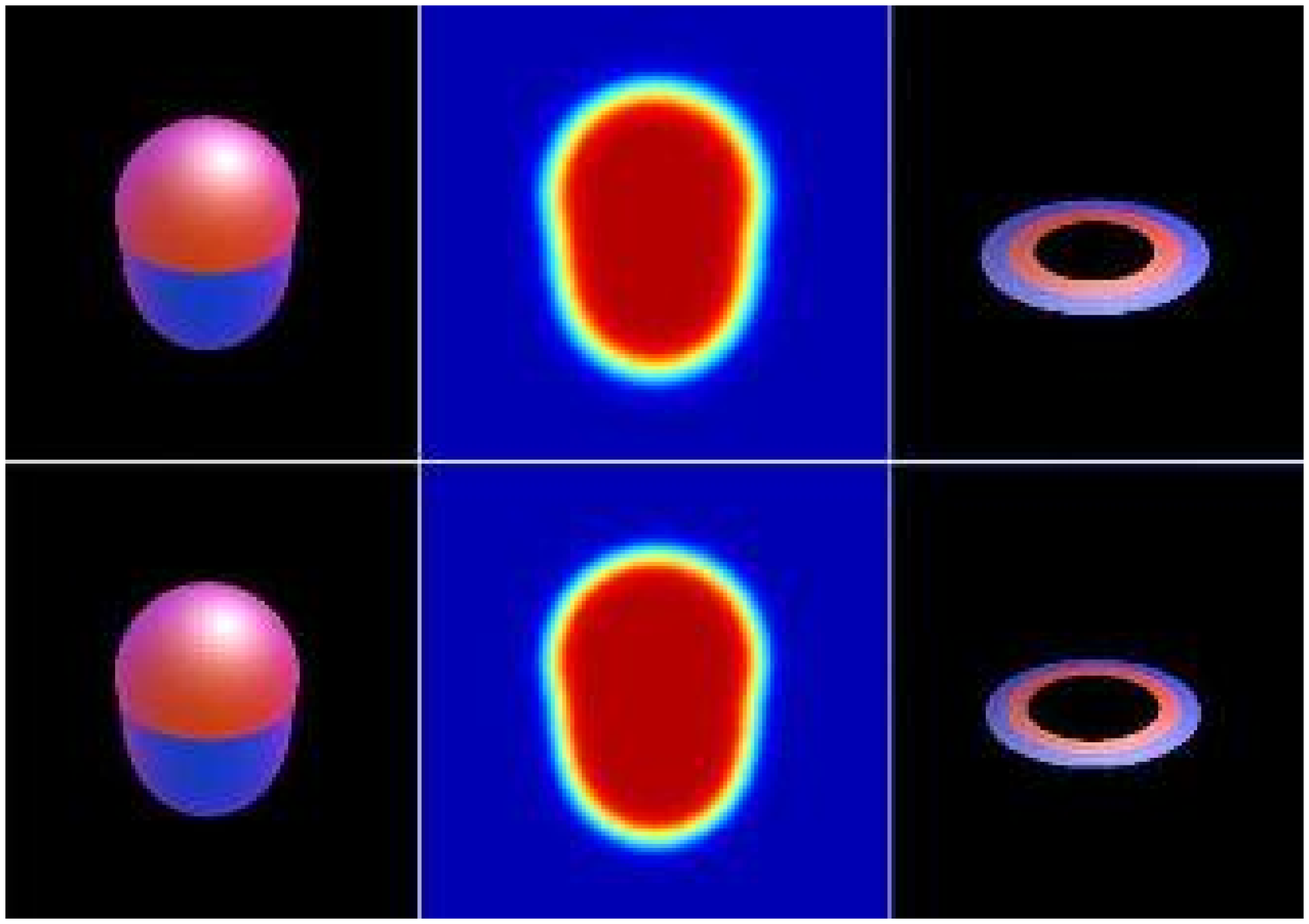}$\quad$
\includegraphics[width=2.0in]{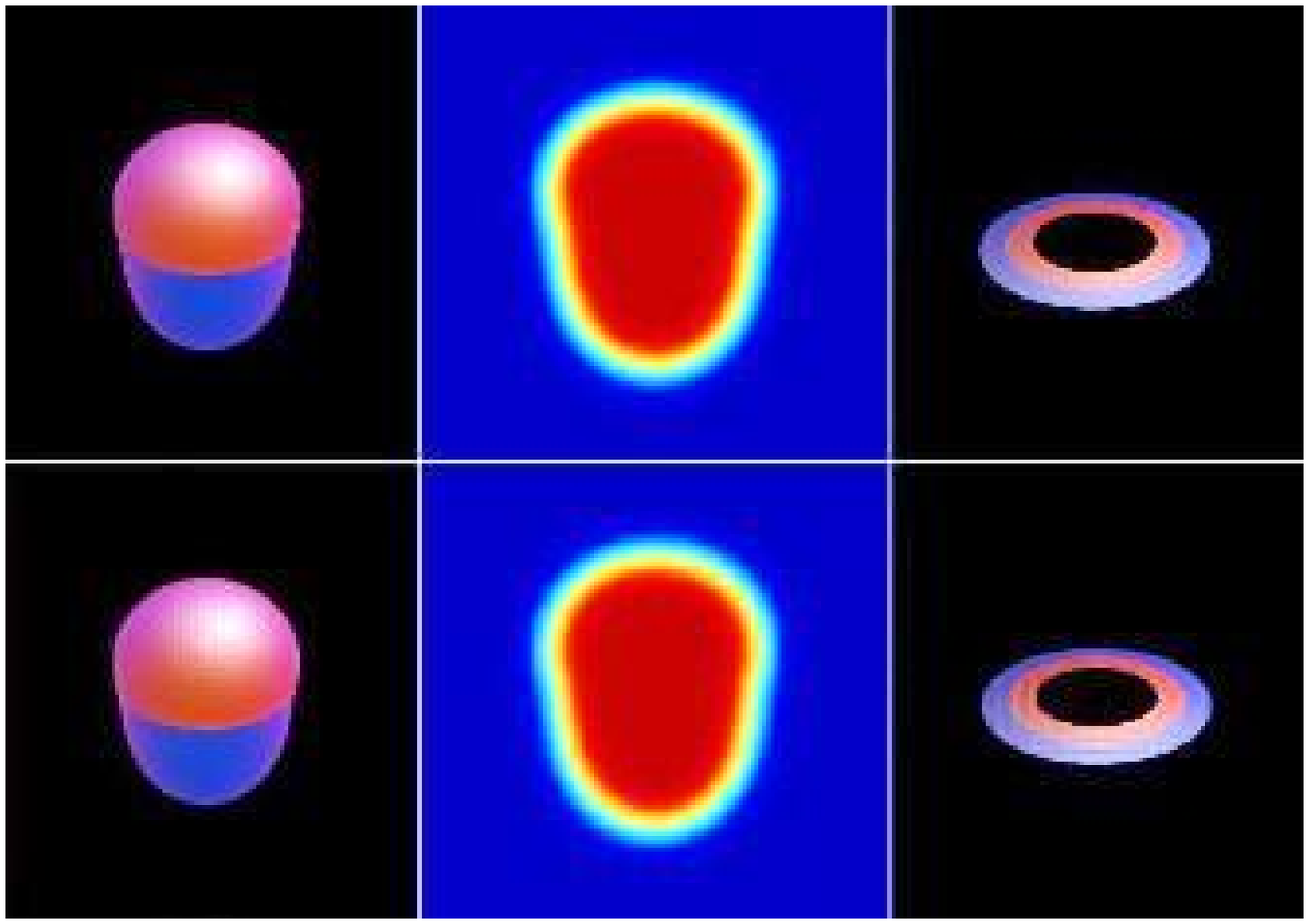}
}
\caption{
The 3d and cross-section views of $\phi$ and the 3d view for part
of $\eta$, computed with different parameters:
left picture, $\epsilon = 2h$ (above) and $\epsilon = 1.5h$ (below);
right picture, 
 $64^3$ grid (above) and $48^3$ grid (below).
} \label{m03}
\end{figure}

The first set of experiments given in Figure \ref{m03} is designed to test
the dependence of the resolution of the phase field function
on the the parameter $\epsilon$ and grid size.
We take a shape similar to the previous
 experiment. First, we take a $64^3$ grid
but use  different values of $\epsilon$ at 
$0.1964(=2h)$ and $0.1472(=1.5h)$. 
The other parameters are defined by $v_d = -216.52$,
$a_0 = 29.46$, $a_d = 0.23$ and $M_i =
 3.2\times 10^5$ for all $i$. The two equilibrium shapes
are almost the same except the transition layer width. The
corresponding final energy values $124.49$ and $123.82$ 
are very closed to each other. The left picture of
Figure \ref{m03} gives the final three dimensional views  and 
some cross section views of the phase field functions $\phi$ and $\eta$.

Now we use the same set of parameters ($\epsilon = 0.1964$, same
initial $\phi_0$ in the same domain), but solve the problem on two
different grid sizes $48^3$ and $64^3$.  We set the parameters 
$v_d = -216.52$, $a_0 = 29.46$,
$a_d = 0.23$ and constants $M_i = 10^4$ for all $i$. The
right picture of Figure \ref{m03} provides the details of the
simulations, with the 3d views of $\phi$ and their density plots
of the cross-sections in $x-z$ plan. The plots of the
corresponding $\eta$ are similar to that in the third column of
the left picture of Figure \ref{m03} and are thus omitted. The
final values of energy are $124.39$ and $124.42$ while the elastic
bending energy values are at $48.05$ and $47.96$, and the line
tension energy values at $76.34$ and $76.46$ respectively. The
close values substantiate the convergence of the simulated
results.

The convergence can also be verified for different penalty
and regularization constants $\{M_i\}_{i=1}^5$. The difference
in the penalty and regularization is to be understood as follows:
the penalty constants $\{M_i\}_{i=1}^3$
are taken to be larger and larger to ascertain the
satisfaction of the volume and areas constraints. The regularization
constants $\{M_4, M_5\}$, on the other hand,
are taken to be smaller and smaller so that while the orthogonality
of the zero level sets of the two phase field functions and the
tanh like profile of $\eta$ are both effectively  maintained in
the simulations, 
the associated energy contributions from the regularization 
terms in fact diminish.

First, we define the Lagrange multipliers by 
 $\lambda_i = \lim_{M_i \rightarrow \infty}
\Lambda_i(\{M_i\}_1^3)$ with $\Lambda_1 = M_1(V(\phi) - v_d)$,
$\Lambda_2 = M_2(A(\phi) - a_0)$, $\Lambda_3 = M_3(D(\phi, \eta)
- a_d)$. 
With other parameters given ($M_4 = M_5 = 10000$, 
$v_d = -216.52$, $a_0 = 29.46$,  $a_d = 0.230$,
$\epsilon = 1.768$, $h = 0.17355$), we set larger and larger
values for $M_1=M_2=M_3$. The results are given in 
Table \ref{tbm} which show that  $\Lambda_1$,
$\Lambda_2$ and $\Lambda_3$ converge to the Lagrange multipliers,
and errors in constraints also decrease. 

\begin{table}
\begin{center}
\begin{tabular}{|c||c|c|c|c|c|}
\hline
 $M_1=M_2=M_3$ &4000 &8000 &16000 & 32000 \\[0.2cm]
  \hline
$\Lambda_1$& -3.0781& -3.0823& -3.0946 & -3.0943 \\[0.2cm]
\hline
$ V(\phi) - v_d $ $(\times 10^{-4})$& -7.6952& -3.8528& -1.9341 & 
-0.9669 \\[0.2cm]
\hline
$\Lambda_2$& 3.6342 & 3.6430 & 3.6608 & 3.6626\\[0.2cm]
\hline
$ A(\phi) - a_0$ $ (\times 10^{-4})$& 9.0855& 4.5537 
& 2.2880 & 1.1445 \\[0.2cm]
\hline
$\Lambda_3$& 0.8144 & 0.8144 & 0.8181 & 0.8173\\[0.2cm]
\hline
$ D(\phi, \eta) - a_d$ $(\times 10^{-4})$& 2.0360 & 1.0180 & 0.5113 
& 0.2554 \\[0.2cm]
\hline
\end{tabular}
\caption{Convergence of the Lagrange multipliers.} 
\label{tbm}
\end{center}
\end{table}

Next, we demonstrate that the regularization terms provide effective 
control on the phase field functions but do not contribute significantly
to the energy minimization.
We set a sequence of decreasing values for $M_4,
M_5$ while taking the same values for $M_1=M_2=M_3=10000$,
and keeping the values of other parameters the same as in the
previous test. The results are given in Table \ref{tbm2}
where $E_4=\frac{1}{2}M_4(N(\phi, \eta))^2$
and $E_5= \frac{1}{2}M_5(P(\eta))^2$ and their ratios with
the total energy $\ce_M$ are provided.
We  can observe the diminishing and
negligible effect of the regularization terms
while there is no noticeable change in the simulated membrane.

\begin{table}
\begin{center}
\begin{tabular}{|c||c|c|c|c|c|}
\hline
$M_4=M_5$ $ (\times 10^{3})$ &32 &16 &8 & 4 \\[0.2cm]
\hline
$E_4$& 0.1223& 0.1108& 0.0966 & 0.0798 \\[0.2cm]
\hline
$E_4/\ce_M$ & 0.0983\%& 0.0892\%& 0.0778\% & 0.0643\% \\[0.2cm]
\hline
$E_5$ & 0.0380& 0.0336& 0.0285 & 0.0227 \\[0.2cm]
\hline
$E_5/\ce_M$ & 0.0305\%& 0.0270\%& 0.0229\% & 0.0182\% \\[0.2cm]
\hline
$\ce_M$ & 124.2942 & 124.2027 & 124.1209 & 124.0500 \\[0.2cm]
\hline
\end{tabular}
\caption{The diminishing effect of regularization on the total energy.}
\label{tbm2}
\end{center}
\end{table}

Having demonstrated the convergence of the numerical algorithms,
we next study the effect of different bending
rigidities and various line tension constants. Then by adjusting
the bending rigidities in the two components and the line tension,
we can simulate the the vesicle shapes in experiment findings
\cite{BHW03}. Unless noted otherwise, the simulation results reported
in the following are obtained with a $64^3$ grid sizes and
$\epsilon = 0.1736$ which can provide sufficient resolution
based on the convergence study.

\begin{figure}[h]
\centerline{
\includegraphics[width=3.0in]{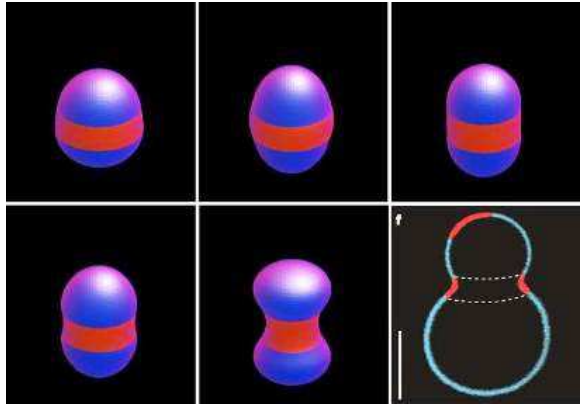}
} \caption{Different values of bending rigidity lead to different
shapes of striped vesicles 
(the right bottom picture is reproduced from \cite{BHW03}).
} \label{m11}
\end{figure}

\paragraph{Effect of the bending rigidities.}
The values of bending rigidities often play a key
role in forming various shapes of vesicles. 
Our first experiment is a simulation of the striped vesicles. We
start from an initial shape where the red
component is situated in the center to give a stripe-looking vesicle. As
shown in the first row of the Figure \ref{m11}, with parameters
$v_d = -213.98$, $a_0 = 29.46$ and $a_d = -13.31$, the initial shape
grows into a very regular stripe-looking ellipsoid shown in the middle of
the first row. In this experiment, the bending rigidity for the
red component is $1.0$ whereas the blue component is $3.0$. With
line tension being fixed at $10.0$, we then make a switch of the bending
rigidity of the two components. As shown in the right picture of
the first row, the red component of the ellipsoid in the middle grows to 
a more cylindrical like shape. Next, 
by preserving the bending rigidity of the blue component while
increasing that of the red component from $3.0$ to $19.0$, the red
component shrinks in the middle and we get a thinner center band
as shown in the left picture of the second row.
It is obvious that the concave region has a smaller mean curvature.
We can further increase the area of the blue component by setting
$a_0 = 33.46$ and $a_d = -19.82$, and with bending rigidities
$3.8$ and $0.2$ respectively for the red and blue components, we get 
the middle picture of the second row in
Figure \ref{m11}. One can compare it with the last picture found
in actual experiments \cite{BHW03} though the differences of
the bending rigidities are not as significant as those used here.
In the final 
shape, the center band has the lowest mean curvature and it is
occupied by the red component (having larger bending rigidities).

As expected, the numerical simulation shows that the component with a
larger bending rigidity is more likely to remain
in regions with smaller values of mean curvature.

\begin{figure}[h]
\centerline{
\includegraphics[width=3.0in]{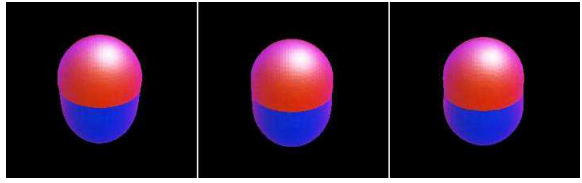}
} \caption{Different values of line tension result different vesicle shapes.}
 \label{m06}
\end{figure}

\begin{figure}[h]
\centerline{\includegraphics[width=3.0in]{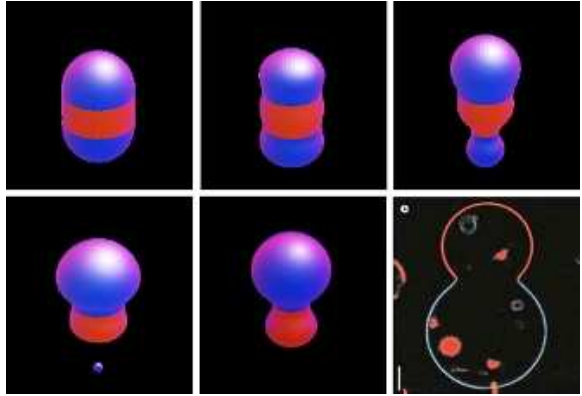}} 
\caption{Effect of line tension: rupture and fission of vesicles components 
(the right bottom picture is reproduced from \cite{BHW03}).} \label{m14}
\end{figure}

\paragraph{Effect of line tension constants.}
By intuition, we expect that larger line tension generally leads to a
shorter interfacial line between two different components. And the line tension
is balanced by the bending and elasticity force and the volume
constraint. In most of the cases, the volume constraint plays a key
role in balancing a large line tension as in the experiments 
illustrated by Figure \ref{m06} and \ref{m14}.

In Figure \ref{m06}, the pictures shown there correspond to 
equilibrium shapes with three
different values of the line tension $10.0, 30.0, 100.0$. The
bending rigidities of the blue colored component is $3.5$ while that 
of the red is $0.5$. By increasing  the line tension, 
the individual components in the two-component vesicle become more
hemisphere like  which are the results of the increasing effect of line 
tension  under the same volume and surface area constraints.

Figure \ref{m14} gives an even more convincing example to the
rupture and vesicle fission observed in this process. As shown in Figure
\ref{m14}, we start from the top left shape. While preserving
$v_d$, $a_0$ and $a_d$ to be $-213.98$, $29.46$ and $-13.31$
respectively, we increase significantly the line tension from
$10.0$ to $100.0$. The vesicle gradually breaks its vertical symmetry
and a small blue vesicle is separated and eventually absorbed
into the top portion through a process like Oswald ripening.
Finally, the vesicle (bottom-right picture of Figure \ref{m14})
only contains two parts, much like the shape observed in the
experiments  \cite{BHW03}.

\paragraph{Comparison with other experimental results.}

We now focus
on the simulations that mimic other two-component vesicle shapes observed in
the experiments of \cite{BHW03}, similar to the 
results depicted in Figures \ref{m11} and \ref{m14}.

\begin{figure}[h]
\centerline{
\includegraphics[width=3.0in]{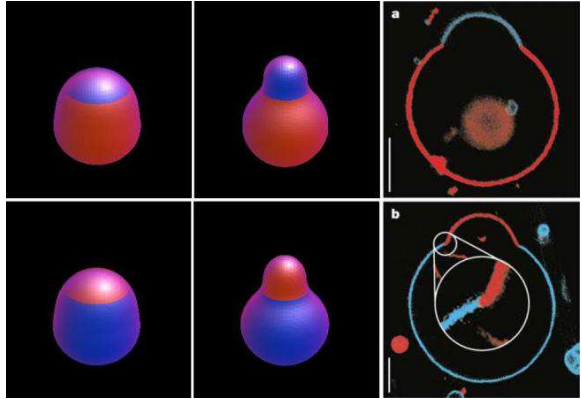}
} \caption{Similar membrane shapes with different areas for
the two components (the pictures on the right column
are reproduced from \cite{BHW03}).} \label{m10}
\end{figure}

As shown  in the two rows of Figure \ref{m10},
we carry out two simulations starting from a shape given on the left.
In both simulations, the red component has bending rigidity $3.0$, and the
blue component has bending rigidity $1.0$. The line tension
between two components is $30.0$. The parameter $v_d$ for volume
constant is $-218.00$, and the surface area parameter $a_0$ is
$29.46$. The parameter $a_d$ giving the difference of surface areas of
the two components takes on the values $18.76$ and $-18.76$ respectively.
The final shapes of the two simulations are shown in the center pictures in both rows. 
One can compare them with the right most experimental picture provided in \cite{BHW03}.

\begin{table}
\begin{center}
\begin{tabular}{|c||c|c|c|c|c|}
\hline
 Energy & $E_r$ & $E_b$ & $E_r+E_b$ & $E_l$ & $E_r+E_b+E_l$ \\
  \hline
  Top  & 58.36 & 12.15  & 70.51 & 138.17 & 208.68 \\
  \hline
  Bottom & 34.83 & 20.06 &  54.89 & 138.01 & 192.90 \\
\hline
\end{tabular}
\caption{Energy comparison for the shapes given in Figure \ref{m10}.}
\label{tbm2a}
\end{center}
\end{table}

The energy values of the two experiments illustrated in Figure
\ref{m10} are given in Table \ref{tbm2a} with energy contributions listed
for individual components and the line tension from the phase boundary. 
We get almost the same line tension energy contribution, but,
as caused by the difference in the bending rigidities,  the
elastic bending energy contributions differ by a factor of
3, which is reflective of the ratio of the bending rigidities.

\begin{figure}[h]
\centerline{
\includegraphics[width=4.0in]{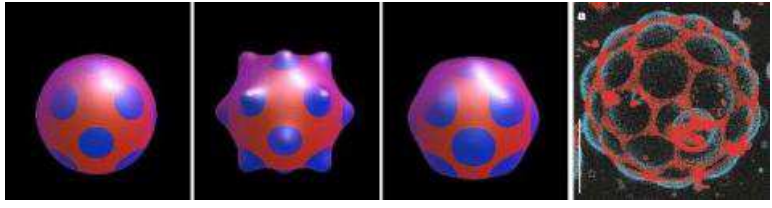}
} \caption{A sphere with disk like bumps: comparing with biological
experiments.} \label{m12}
\end{figure}

\begin{figure}[h]
\centerline{
\includegraphics[width=3.0in]{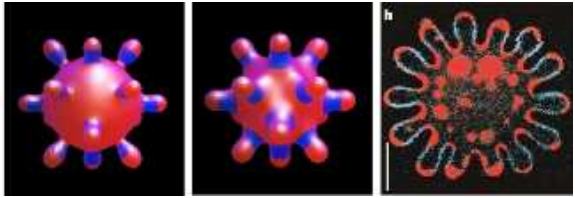}
} \caption{Two-component shape with 14 bumps
(the last picture is from \cite{BHW03}).} \label{m13b}
\end{figure}

We now turn to simulate a couple of other interesting shapes experimentally
observed in \cite{BHW03}  as illustrated in the last pictures of Figure \ref{m12} and 
Figure \ref{m13b} respectively. In Figure \ref{m12}, we
first start from a spherical surface which is divided into two components 
where one component occupies similar spherical caps in  twelve well-spaced  locations
on the membrane surface. The bending rigidity is $3.5$ for the red component 
and $0.5$ for the blue component, and
the line tension is $10.0$. With a larger surface area of the blue
component and a smaller volume than those values for the exact
sphere, the blue component (with smaller bending rigidity)
starts to bulge. The parameters are taken respectively as
$v_d = -174.17$, $a_0 =54.63$ and  $a_d = 11.01$. If we further 
increase the volume and enlarge the relative
area of the blue component by increasing $v_d$ to $
-167.00$, while keeping $a_0$ at $54.63$  and changing $a_d$ to
$5.00$, the resulting computed shape (the third picture in
Figure \ref{m12}) is very similar to the experiment
findings \cite{BHW03} (the last picture in Figure \ref{m12}).

\begin{figure}[h]
\centerline{
\includegraphics[width=3.0in]{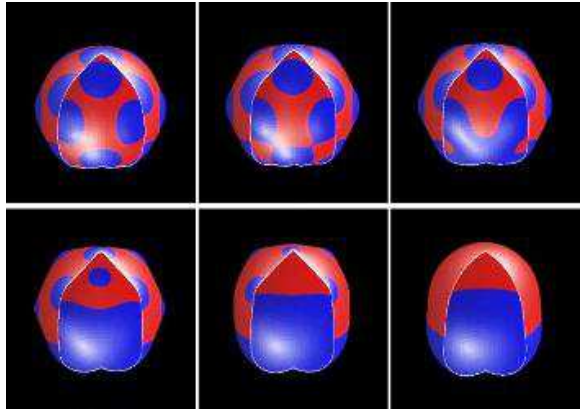}
} \caption{The merging of blue component (cut view).} \label{m12m}
\end{figure}

The shape corresponding to the third picture of Figure \ref{m12} stays 
as a near equilibrium (meta-stable) 
state for a range of parameter values. But if we further increase
the area of the blue component, for example, by setting $a_d =
2.5$,  further coarsening of the blue components will take
place. The merger of disconnected components continues, much like
the Oswald ripening effect, and eventually transforms into shapes
similar to that presented earlier in Figures \ref{m03} and
\ref{m06}. The transformation is illustrated in Figure \ref{m12m}.

\begin{figure}[h]
\centerline{
\includegraphics[width=3.0in]{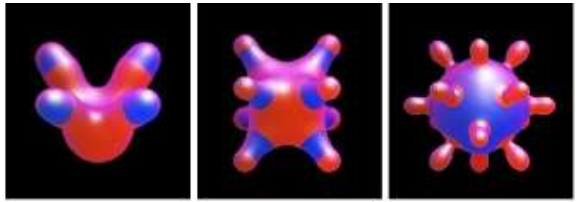}
} \caption{Various shapes of  two component membranes.} \label{m13}
\end{figure}

Next,  we take an initial membrane profile similar to that in the
second picture of Figure \ref{m12}. By setting $v_d = -203.00$,
$a_0 = 54.63$ and $a_d = 5.00$ so that both the total volume and
the area of the red components are decreased, we can then observe
the growth of bumps of the red component, leading to a shape shown
in the right pictures of Figure \ref{m13}. Take other initial
profiles, other equilibrium shapes as shown in the
left and center pictures in Figure \ref{m13} have also been observed
in our simulations.

From Figure \ref{m13}, it can be seen that the two-component
vesicles may display very rich patterns, even in the
absence of spontaneous curvature effect. One naturally may wonder
if some of them are experimentally observable. The next
set of experiments
draws inspiration from the center and right figures of Figure
\ref{m13} and leads to interesting comparisons with similar experimental
observations in \cite{BHW03}. We start with the same phase field
$\phi$ as the profile in the right picture of Figure \ref{m13},
but use a modified $\eta$ such that the neck of the bumps are formed
by the blue component as the case of the center picture of Figure
\ref{m13}. This leads to  an initial shape as shown in the left
picture of Figure \ref{m13b}. Setting the parameters as $v_d =
-203.00$, $a_0 = 54.63$, and $a_d = 22.68$, we finally get a
shape (center picture of Figure \ref{m13b}) very close to the
experimentally observed shape given in \cite{BHW03}  (right picture of Figure
\ref{m13b}).

Shapes depicted in \ref{m13b} are fairly robust. In fact,
with a slight modification of the final shape and a rotation with 
a given angle, then we found that using the gradient flow,
the equilibrium solution is again in the shape (except
for a rotation). Results of such calculations on both
$64^3$ and $96^3$ grids are given for comparison.

 \begin{figure}[h]
\centerline{
\includegraphics[width=2.0in]{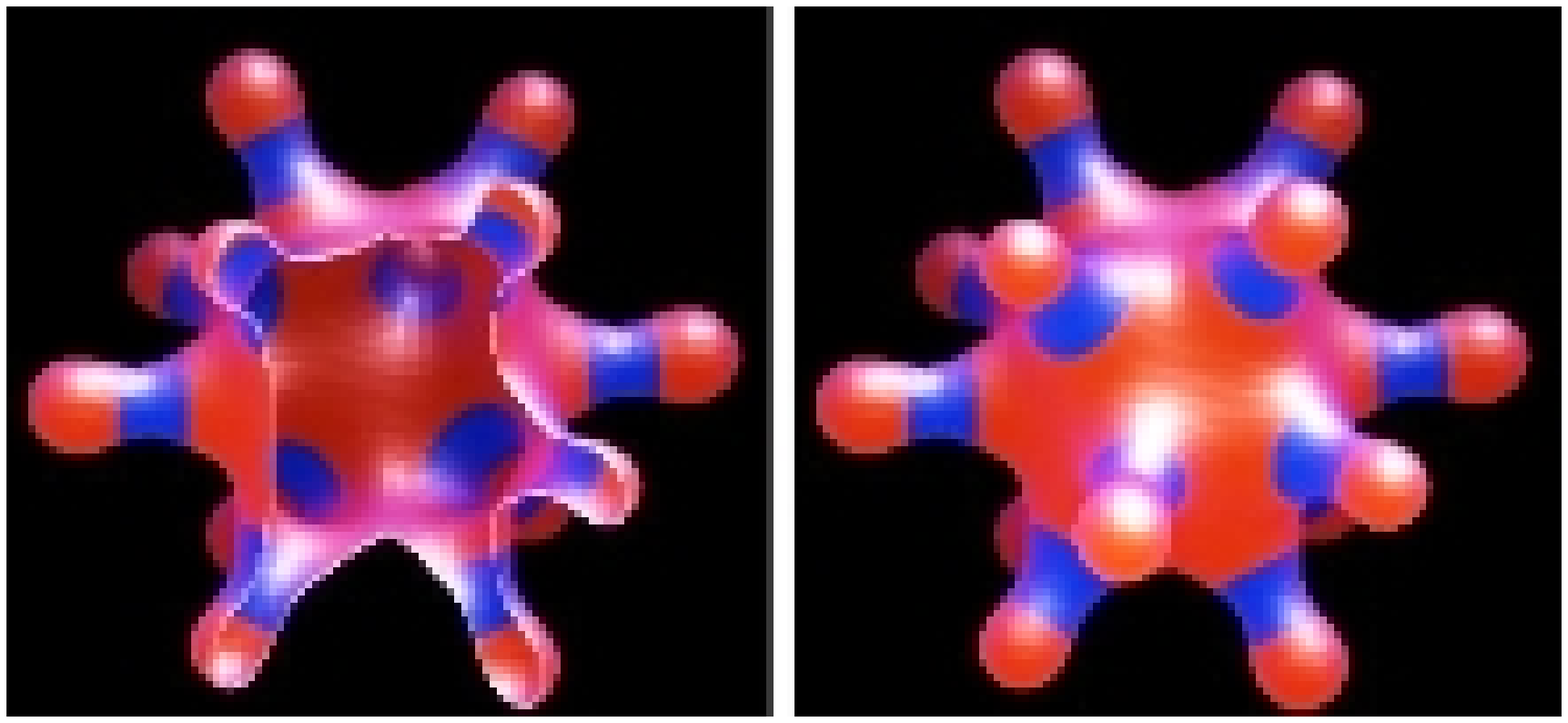}
\includegraphics[width=2.0in]{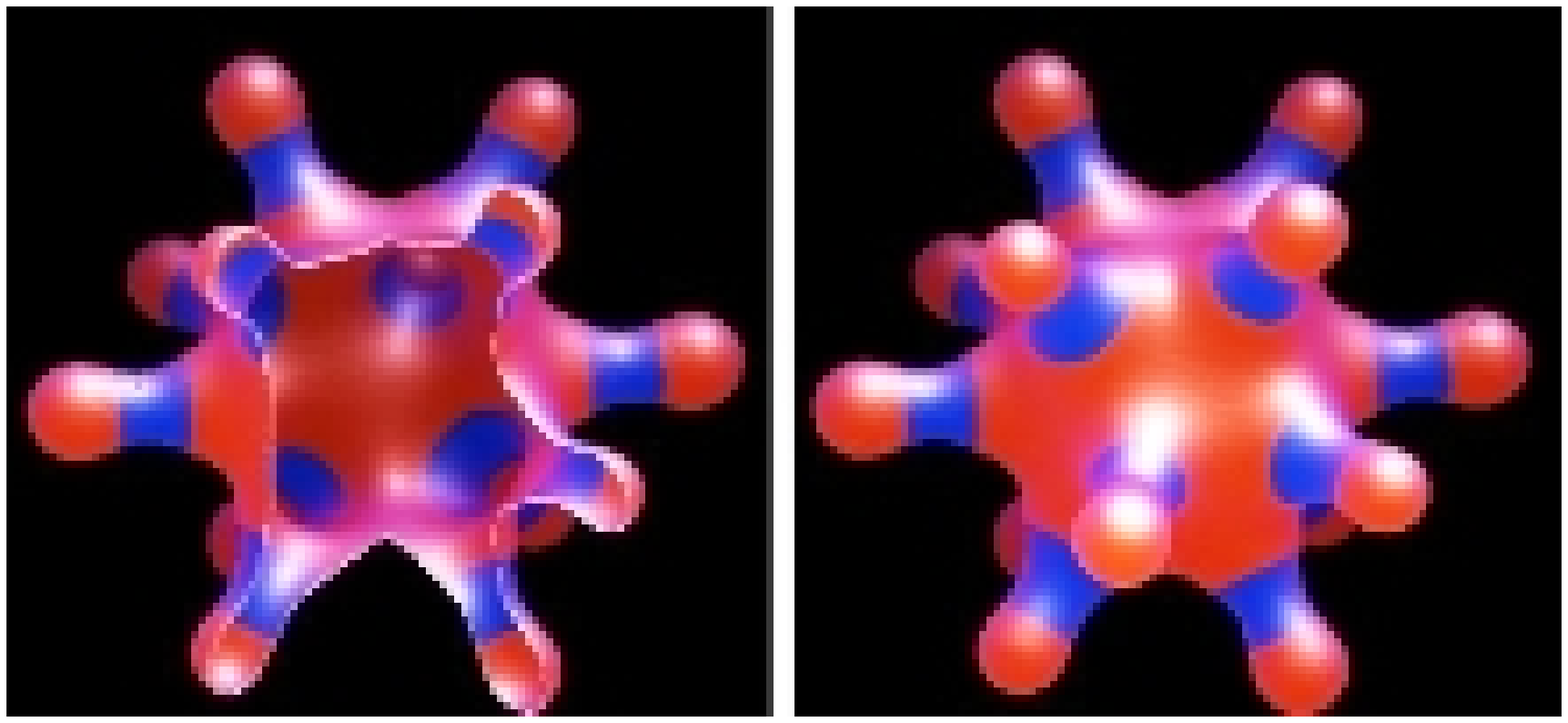}
} \caption{Rotated two-component shapes with 14 bumps
computed by different meshes.} \label{m14r}
\end{figure}

\section{Open liposomal membranes}

In this section, we apply similar ideas to model open lipid
membranes. The transformations
from vesicles to open membranes and the reverse process from open
membranes to vesicles were first observed in \cite{SaTaHo98}.
Here, we only consider the one-component open membranes with
specified surface areas. The total energy of
an open membrane $\Gamma$ with edge $\gamma_0$
 may be conveniently defined as the
sum of the elastic bending
energy and the line tension energy \cite{cgs02,TuOu03,TuOu04,umeda,yyn05}:
$$
 \int_\Gamma \left( a_1 + a_2 (H-c_0)^2 + a_3 G \right)\, ds +
\int_{\gamma_0} \delta \, dl\;.
$$
For simplicity, we set the surface tension $a_1$
and the line tension $\delta$, as two constants, we also do not
consider the contribution of the geodesic curvature term in the line
energy on the boundary.  The effects of  surface tension, the Gaussian
and spontaneous curvatures are also ignored. Our problem is then to minimize 
the following total energy
$$
E_o = \int_\Gamma k H^2 \, ds + \int_{\gamma_0} \delta \, dl
$$
with prescribed surface area $|\Gamma|$.

Most of the available numerical simulations for open
membranes have largely been confined to axis-symmetric cases
based on the variational calculation of the above energy.
We hereby develop a new phase field model for open
membranes, and present some numerical simulations
for the full three dimensional case to demonstrate the
effectiveness of the model.

\paragraph{Phase field model for open membranes.}
We can treat open membranes as two component membranes with one
component having zero bending rigidity. Again, we let
$\gamma_0$ be the intersection of two
orthogonal surfaces $\Gamma$ and $\Gamma_0$ which are
implicitly defined as the level-set of the functions $\phi$ and $\eta$
respectively.

Now we denote $c(\eta) = \frac{1}{2}(1 + \tanh(\frac{\eta}{\xi}))$,
and let the line tension energy be still formulated by $L(\phi, \eta)$ 
in (\ref{eml}), with the  elastic bending energy of the membrane 
$$E(\phi, \eta) =
\int_{\Omega} \frac{k c(\eta)}{2\epsilon}
\big(\epsilon\Delta\phi + \frac{1}{\epsilon} \phi
(1-\phi^2)\big)^2 \, dx,$$
Then, our phase field model for open membranes is to minimize
$E(\phi, \eta) + L(\phi, \eta)$
with the surface area constraint
\begin{equation}
D(\phi, \eta) = \int_\Omega
c(\eta)[\frac{\epsilon}{2}|\nabla\phi|^2 +
\frac{1}{4\epsilon}(\phi^2 -1)^2] \, dx = a_d\;.
\end{equation}
Similar to the two-component vesicle case studied earlier,
to maintain the good profiles for both phase field functions
$\phi$ and $\eta$ and the orthogonality of $\Gamma$ and
$\Gamma_\perp$, we can again take the penalty formulation
\begin{eqnarray}
\ce_M(\phi, \eta) &=& W(\phi, \eta) + L(\phi, \eta) +
\frac{1}{2}M_3(D(\phi, \eta) - a_d)^2 \nonumber\\&&+
\frac{1}{2}M_4(N(\phi, \eta))^2 + \frac{1}{2}M_5(P(\eta))^2 +
\frac{1}{2}M_6(P(\phi))^2\;.\label{oengy}
\end{eqnarray}
Then, we can again use a gradient flow like (\ref{eq3d})
to compute the equilibrium shapes by a similar  numerical scheme as 
that given in section \ref{sec:mns}.

\paragraph{Numerical simulations of open membranes}
We now present some numerical simulations of open
membranes and compare them with biological experimental
findings. Most of the model and simulation
parameters are chosen to be in the same range  as that for the two component
vesicle simulations in the earlier section.

\begin{figure}[h]
\centerline{
\includegraphics[width=4.0in]{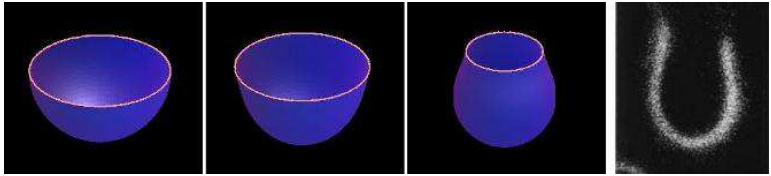}
} \caption{Open membranes with different line tensions
(the right most picture is reproduced from \cite{SaTaHo98}).} \label{m31}
\end{figure}
Figure \ref{m31} gives the simulation results of a simple open membrane.
Starting from a half sphere (the left picture), with bending
rigidity $k = 1.0$ and line tension $\delta = 1.0$, we get an
equilibrium shape shown in the second picture. If a larger
line tension $\delta = 1.28$ is used,  an equilibrium shape is reached
as that in the third picture. One can compare it with the right picture
obtained in the biological experiments described in \cite{SaTaHo98}.  
We note that  the elastic bending energy are $10.12$ and $15.96$
and the line tension energy are $10.94$ and
$7.95$ respectively for the solutions in the  second and third pictures.

\begin{figure}[h]
\centerline{
\includegraphics[width=4.in]{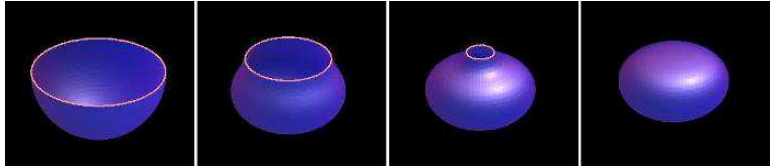}
} \caption{Open membrane closes due to large line tension.}
\label{m32}
\end{figure}

The time evolution snapshots are given in  Figure \ref{m32} where
the line tension is taken as $\delta = 25.0$.
The simulation results show that, when the line
tension becomes large enough, the open membrane becomes self-enclosed.

\begin{figure}[h]
\centerline{
\includegraphics[width=4.0in]{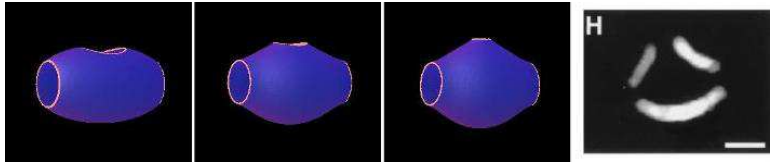}
} \caption{Open membranes with three holes
(the right most picture is reproduced from \cite{SaTaHo98}).} \label{m33}
\end{figure}

Finally, in Figure \ref{m33}, we simulate a shape (the right picture) with
three holes as observed in an experiment of \cite{SaTaHo98}. Starting from
the left most picture corresponding to an ellipsoid with three holes, 
setting the bending rigidity $k = 1.0$ and line tension $\delta = 1.0$, 
and following the gradient
flow of the energy, the initial shape starts to deform first into an intermediate
shape given in the second picture. The computed
equilibrium shape is shown in the third picture which again shows striking
similar to the experimental finding.

\section{Conclusion}

In this paper, we formulated a phase field model for the multi-component
vesicles membranes, and as a special case, the open membranes
with free edges. The models incorporate the effect of the
elastic bending energy together with the line tension between each
two components. Full three dimensional numerical simulations
presented here demonstrate that the experimental observations 
given in  \cite{BHW03} 
can be effectively simulated by the phase field bending elasticity
and line tension model. Furthermore, the simulation results illustrate
that many experimentally observed exotic patterns such as bud formation
and vesicle fission can appear in two-component vesicles 
due to the inhomogeneous bending stiffness and
the competition of the bending energy and the interfacial 
line tension even without incorporating the spontaneous curvature
or the asymmetry of the bilayer.

In conclusion, we point out the this
generalization of our diffusive interface model to
two component vesicle membranes fits nicely into the
previously established unified framework for the derivation
of dynamic and static equations and the development of
numerical algorithms and codes. Many issues remain to be
examined in future works. First,
there maybe other more effective ways in formulating the line tension
energy, including the use of geodesic curvature along the phase boundary,
which may be important to model the difference of the stretching rigidity in
two components \cite{BSWJ05}.
Second, more rigorous analysis of our models are needed in the
future. Third, in our numerical simulations, we have not examined the
effect of the spontaneous curvature as we have done for the
one component case \cite{DuLiRoWa042}. It is expected that
more complex shapes would be be discovered in this case.
Moreover, the interaction of multi-component vesicles with the fluid
and electric fields are also exciting topics to be investigated
further in the future.

\section*{Appendix: Justification of the energy and constraints.}
 We now provide some brief calculations to  rationalize
the definitions of the energy functional and the constraints in
the phase field setting. Same as the discussion in
\cite{DuLiWa03}, we can first illustrate that in a very broad
ansatz, for small $\epsilon$ and $\xi$, minimizing $W(\phi, \eta)$
leads to a phase field function $\phi(x)$ which is approaching to
$\tanh(d(x, \Gamma)/(\sqrt 2 \epsilon))$ as $\epsilon \rightarrow
0$. In fact for small $\epsilon$, due to the uniform bound of the
functional $B$, the region away from the level set $\phi = 0$ are
all close to $\phi = +1$ or $-1$. In such cases, one  can always
define the following transformation near the interface:
\begin{equation}\label{phi}
\phi (x) = q^\epsilon (\frac{d(x)}{\epsilon}),
\end{equation}
where $d(x)$ is the distance of the point $x\in \Omega$ to the
surface $\Gamma$.  Substituting this into (\ref{eme}), we have
that:
\begin{equation}\label{ener}
E(\phi) = \int_\Omega  \frac{k(\eta)}{2 \epsilon}
\left|{q^\epsilon}' (\frac{d(x)}{\epsilon}) \Delta d(x) +
\frac{1}{\epsilon}({q^\epsilon }'' -({q^\epsilon}^2 -1) q^\epsilon
)
\right|^2 \, dx.
\end{equation}
If we keep $k(\eta)$ positive, as $\epsilon \rightarrow 0$,  to minimize the
energy, the leading term in the above has to vanish, that is,
\begin{equation}
\left|{q^\epsilon }'' -({q^\epsilon}^2 -1) q^\epsilon\right|^2
\rightarrow 0
\end{equation}
which means that the transition region  profile $q^\epsilon (\cdot)
$ is approaching to the function $\tanh ({\frac{\cdot}{\sqrt 2}})$.
In the meantime, we see that $\phi$ is approaching to the Heaviside
function with $1$ inside of the interface and $-1$ outside. $\Gamma$
still coincides with the zero level set of $\phi$. Moreover
(\ref{phi}) indicates that the parameter $\epsilon$ is effectively
the thickness of the transition region between $\{\phi =1\}$ and
$\{\phi = -1\}$. One can refer \cite{DuLiRoWa041} for more rigorous
proof of this.

Now, we denote $s(\phi) = \frac{\epsilon}{2}|\nabla\phi|^2 +
\frac{1}{4\epsilon}(\phi^2 -1)^2$. When the line tension $L(\phi,
\eta)$ reach its minimum, we have 
$$
\frac{\delta L}{\delta \eta} = f(\eta, s) = -\epsilon \nabla \cdot
(s \nabla \eta) + \frac{1}{\epsilon}s(\eta^2 -1)\eta = 0.
$$
For $\phi = \tanh(d/(\sqrt 2 \epsilon))$, $s =
\frac{1}{2\epsilon}(\phi^2 -1)^2$ and therefore $\nabla s \cdot
\nabla \eta = 0$ as $\nabla \phi \cdot \nabla \eta = 0$. Then
$f(\eta, s) = 0$ means
$$
-\epsilon \Delta \eta + \frac{1}{\epsilon}(\eta^2 -1)\eta = 0\;.
$$
If we write $\eta$ again by $q^\epsilon(d(x,
\Gamma_\bot)/\epsilon)$, from the above equation we have
$$
- {q^\epsilon}'(\frac{d}{\epsilon})\Delta d(x) + \frac{1}{\epsilon}
\big(({q^\epsilon}^2 - 1)q^\epsilon - {q^\epsilon}'' \big) = 0.
$$
As $\epsilon \rightarrow 0$, we have $({q^\epsilon}^2 -
1)q^\epsilon - {q^\epsilon}'' = 0$. To minimize $L$, we can expect
that far away from the $\Gamma_\bot$, $q^\epsilon$ is $+1$ or
$-1$, therefore we also have $q^\epsilon(x)= \tanh(\frac{x}{\sqrt 2
\epsilon})$. On the other hand, we can use the same argument for
$\phi$ if we know $\eta$ is a tanh function, which would further
strengthen the ansatz that $\phi$ and $\eta$ are both tanh functions 
to lead order of $\epsilon$. In fact, following more careful analysis
as those in \cite{DuLiRoWa041}, we expect that the differences between
$\phi$ and $\eta$ and the respective
tanh profiles are second order in $\epsilon$
which would allow us to rigorous derive the asymptotic limits
(\ref{eqsharp1}-\ref{eqsharp3}).

\section*{Acknowledgment}
The experimental pictures used in the various
figures of this paper are from a couple of sources
with permission from the authors:
the two component membranes experiments are from
\cite{BHW03}, and the open membranes are from
\cite{SaTaHo98}.

\def\cprime{$'$}


\begin{thebibliography}{99}

\bibitem{AMW98}
{\sc D. Anderson, G. McFadden and A. Wheeler}, 
{\em Diffusive-interface methods in fluid mechanisms}, 
Ann. Rev. Fluid Mech. 30, pp.~139-165,  1998. 

\bibitem{BHW03}
{\sc T.~Baumgart, S.~Hess and W.~Webb}, {\em Imaging coexisting
fluid domains in biomembrane models coupling curvature and line
tension}, Nature, Vol 425, pp.~821--824, 2003.

\bibitem{BSWJ05}
{\sc T.~Baumgart, S.~Hess, W.~Webb and T. Jenkin},
{\em Membrane Elasticity in Giant Vesicles with Fluid Phase Coexistence},
Biophysical Journal, 89, pp.1067-1080, 2005.

\bibitem{BeMc92}
{\sc D.~Benvegnu and M.~McConnell}, {\em Line tension between liquid
domains in lipid monolayers}, Journal of Physical Chemistry, 96,
pp.~6820-6824, 1992.

\bibitem{misbah}
{\sc T. Biben, K. Kassner and C. Misbah}, {\em Phase-field
approach to 3D vesicle dynamics},  Phys. Rev. E., 72, pp.041921,
2005.

\bibitem{BWBK02}
{\sc 
W. Boettinger, J. Warren, C. Beckermann, and A. Karma}, 
{\em  Phase-field simulation of solidification}, 
Annual Review of Materials Research, 32, pp.163-194, 2002

\bibitem{osher01}
{\sc P. Burchard,  L.-T. Cheng, B. Merriman and S. Osher}, {\em
Motion of Curves in Three Spatial Dimensions Using a Level Set
Approach}, Journal of Computational Physics, 170, pp.720-741,
2001.

\bibitem{cgs02}
{\sc R. Capovilla, J. Guven and J. Santiago},
{\em Lipid membranes with an edge},
Phys. Rev. E, 66, pp.021607, 2002.

\bibitem{CaCh92}
{\sc G.~Caginalp and X.~F. Chen}, {\em Phase field equations in the
singular limit of sharp interface problems}, in {\em On the
evolution of phase boundaries}
 (Minneapolis, MN, 1990--91), Springer, New York, 1992, pp.~1--27.


\bibitem{Ch02}
{\sc L.-Q. Chen}, {\em Phase-field models for microstructure evolution},
{ Annual Review of Materials Research},
{32}, pp.~113-140, 2002.  

\bibitem{Ci98}
{\sc P.~G. Ciarlet}, {\em Introduction to linear shell theory},
V.~1 of
  Series in Applied Mathematics (Paris), Gauthier-Villars, \'Editions
  Scientifiques et M\'edicales Elsevier, Paris, 1998.

\bibitem{Ci00}
\leavevmode\vrule height 2pt depth -1.6pt width 23pt, {\em
Mathematical
  elasticity. {V}.{III}}, V.~29 of Studies in Mathematics and its
  Applications, North-Holland Publishing Co., Amsterdam, 2000.
\newblock Theory of shells.


\bibitem{DoKaNoSp93}
{\sc H.~D\"{o}bereiner, J.~K\"{a}s, D.~Noppl, I.~Sprenger and
E.~Sackmann}, {\em Budding and fission of vesicles}, Biophysical
Journal, 65, pp. 1396¨C1403, 1993.

\bibitem{DuLiRoWa041}
{\sc Q.~Du, C.~Liu, R.~Ryham and X.~Wang}, {\em A phase field
formulation of the Willmore problem}, Nonlinearity, 18, pp.
1249-1267, 2005.

\bibitem{DuLiRoWa042}
{\sc Q.~Du, C.~Liu, R.~Ryham and X.~Wang}, {\em Modeling the
Spontaneous Curvature Effects in Static Cell Membrane Deformations
by a Phase Field Formulation}, Communications in Pure and Applied
Analysis, 4, pp. 537-548, 2005.

\bibitem{DuLiRyWa06}
{\sc Q.~Du, C.~Liu, R.~Ryham and X.~Wang}, {\em Modeling Vesicle
Deformations in Flow Fields
 via Energetic Variational Approaches}, preprint, 2006.

\bibitem{DuLiWa03}
{\sc Q.~Du, C.~Liu, and X.~Wang}, {\em A phase field approach in the
numerical study of the elastic bending energy for vesicle
membranes}, Journal of Computational Physics, 198, pp. 450-468,
2004.

\bibitem{DuLiWa04}
{\sc Q.~Du, C.~Liu, and X.~Wang}, {\em Retrieving Topological
Information For Phase Field Models}, SIAM Journal on Applied
Mathematics 65, pp. 1913-1932, 2005.

\bibitem{DuLiWa06}
{\sc Q.~Du, C.~Liu, and X.~Wang}, {\em Simulating the Deformation of
Vesicle Membranes under Elastic Bending Energy in Three Dimensions},
Journal of Computational Physics, 212, pp. 757-777, 2006.

\bibitem{gg98}
{\sc W. Gozdz and G. Gompper},
{\em Shapes and shape transformations of two-component membranes of
 complex topology},
Phys. Rev. E 59, 4305-4316 (1999).

\bibitem{jiang00}
{\sc  Y. Jiang,  T. Lookman, and A. Saxena}, {\em Phase separation and
shape deformation of two-phase membranes}, Phys. Rev. E. 6 (2000),
R57-R60.

\bibitem{lip96}
{\sc F. Juelicher and R. Lipowsky}, {\em Shape transformations of
vesicles with intramembrane domains}, Phys. Rev. E. 53, 2670-2683, 1996.

\bibitem{kgl01}
{\sc P. Kumar, G. Gompper and R. Lipowsky},
{\em Budding Dynamics of Multicomponent Membranes},
 Physical Review Letters, 
86, pp.3911-3914 , 2001

\bibitem{Li92}
{\sc R.~Lipowsky}, {\em Budding of membranes induced by
intramembrane domains}, Journal de Physique II, France 2,
pp.~1825-1840, 1992.

\bibitem{Li95}
{\sc R.~Lipowsky}, {\em The morphology of lipid membranes}, Current
Opinion in Structural Biology, 5, pp.~531--540, 1995.


\bibitem{Li02}
{\sc R.~Lipowsky}, {\em Domains and Rafts in Membranes 
 Hidden Dimensions of Self-organization},
Journal of Biological Physics, 28, pp.195-210, 2002.
 
\bibitem{MAV04}
{\sc J. McWhirter, G. Ayton and G. Voth},
{\em Coupling Field Theory with Mesoscopic Dynamical Simulations of Multicomponent Lipid Bilayers}, Biophysical Journal 87, pp.~3242-3263, 2004.
 
\bibitem{OuLiXi99}
{\sc Z.~Ou-Yang, J.~Liu, and Y.~Xie}, { Geometric Methods in
the Elastic Theory of Membranes in Liquid Crystal Phases}, World
Scientific, Singapore, 1999.

\bibitem{OR02}
{\sc S.~Osher and R.~Fedkiw}, {The Level Set Method and Dynamic
Implicit Surfaces}, Springer-Verlag, 2002.

\bibitem{SaTaHo98}
{\sc A.~Saitoh, K.~Takiguchi, Y.~Tanaka and H.~Hotani}, {\em
Opening-up of liposomal membranes by talin}, Proceedings of the
National Academy of Sciences, Biophysics, 956, pp.~1026-1031, 1998.

\bibitem{Se92}
{\sc U.~Seifert}, {\em Curvature-induced lateral phase separation in
two-component vesicles}, Physical Review Letters, 70,
pp.~1335-1338, 1993.

\bibitem{Se99}
{\sc J.~A. Sethian.}
 Level Set Methods and Fast Marching Methods: evolving interfaces
 in computational geometry, fluid mechanics, computer vision, and materials
  science.
 Cambridge University Press, New York, 2nd edition, 1999.

\bibitem{TuOu03}
{\sc Z.~Tu and Z.~Ou-Yang}, {\em Lipid membranes with free edges},
Physical Review E, 68, 061915 (1-7), 2003.

\bibitem{TuOu04}
{\sc Z.~Tu and Z.~Ou-Yang}, {\em A geometric theory on the
elasticity of bio-membranes}, Journal of Physics A: Mathematical and
General, 37, pp.~11407-11429, 2004.

\bibitem{umeda}
{\sc T. Umeda, Y. Suezaki, K. Takiguchi, and H. Hotani},
{\em Theoretical analysis of opening-up vesicles with single and two holes},
Phys. Rev. E, 71, pp.011913 (1-8), 2005.

\bibitem{wangthesis}
{\sc X.~Wang}, {\em Phase Field Models and Simulations of Vesicle
Bio-membranes}, Ph.D thesis, Department of Mathematics,
 Penn State University, 2005.

\bibitem{yyn05}
{\sc Y. Yin, J. Yin and D. Ni},
{\em General Mathematical Frame for Open or Closed Biomembranes I: Equilibrium Theory and Geometrically Constraint Equation}, 
Journal of Mathematical Biology, 51, pp.~403-413. 

\end{thebibliography}
\end{document}